\documentclass[aps,prb,reprint,groupedaddress]{revtex4-1}
\usepackage{graphicx}
\usepackage{multirow}
\usepackage{mathrsfs}
\usepackage{amsfonts}
\usepackage{amssymb}

\newcommand{\e}{\epsilon }
\begin{document}


\title{Non-Markovian coherent feedback control of quantum dot systems}


\author{Shibei Xue$^{1,2,3}$}\email[]{xueshibei@gmail.com}
\author{Re-Bing Wu$^{1,2}$}\email[]{rbwu@tsinghua.edu.cn}
\author{Michael R. Hush$^{3}$}
\author{Tzyh-Jong Tarn$^{1,2,4}$}
\affiliation{$^1$Department of Automation, Tsinghua University, Beijing 100084, P. R. China\\
$^2$Center for Quantum Information Science and Technology, TNList, Beijing 100084,
P. R. China\\
$^3$School of Information Technology and Electrical Engineering, University of New South Wales Canberra at the Australian Defence Force Academy, Canberra, ACT 2600, Australia\\
$^4$Department of Electrical and Systems Engineering, Washington University, St. Louis, Missouri 63130, USA}

\date{\today}

\begin{abstract}
This paper presents a non-Markovian coherent feedback scheme to control single quantum dot systems. The feedback loop is closed via
a quantum tunneling junction between the natural source and drain baths of the quantum dot. The exact feedback-controlled non-Markovian Langevin equation is derived for describing the dynamics of the quantum dot. To deal with the nonlinear memory function in the Langevin equation, we analyze the Green's function-based root locus, from which we show that the decoherence of the quantum dot can be suppressed via increasing the feedback coupling strength. This effectiveness of decoherence suppression induced by non-Markovian coherent feedback is verified by an example of single quantum dot systems.
\end{abstract}


\maketitle

\section{Introduction}

As a solid-state information carrier for quantum computation, quantum dot systems have attracted much attention in recent years~\cite{LossPRA1998,BurkardPRB1999,ElzerNAT2004,KrouNAT2004}. As well as other quantum registers, the coherent manipulation of the quantum dot is vital for processing quantum information~\cite{Mansci2015,franco}, which is always deteriorated by the decoherence induced by interaction with the environments~\cite{LeeJCP2008,PhysRevA.89.042320,PhysRevA.77.032117}. In quantum dot systems, the interaction occurs between the quantum dot and source and drain electrodes, the hyperfine
interaction between electron spins of quantum dots and spins of nuclei, and the noise generated by the defects on the substrate materials~\cite{Chirolli2008,Tu2008,Xue2011}.


Under conditions that the memory time is ignorable, the Markovian approximation can be taken to simplify the analysis and design of open quantum control systems such as stabilizing the current through nanostructures and purifying the state of quantum dot qubit via feedback control~\cite{TobiasPRL2010,PoltlPRB2011,BluPRL2010}.
However, for general cases, the feedback control performance may be degraded due to the violation of Markovian approximation in the solid-state systems. Consequently, colored noise disturbs the system of interest, whose spectrum is defined by a multiplication of the state density and the square norm of the coupling strength between the system and the environment~\cite{leggett1987}.
This resulting non-Markovian effect can be harnessed by using a class of direct coherent feedback approach~\cite{XuePRA2012} where the structure of the environments is altered by the couplings between the modes of the environment, and the characteristics of correlated environments can modify the non-Markovianity of a quantum system~\cite{PhysRevA.92.012315,zhuEPJD}.
In particular, no measurement on the non-Markovian dynamics is required with this method.

This paper studies coherent feedback control of non-Markovian dymanics of single quantum dot systems, with application to the suppression of decoherence where the noise baths, i.e., source and drain, are coupled together by a tunneling junction to form a closed loop. The quantum transport process of the electrons between them is modified via adjusting the structure of the junction so that the effective noise spectrum of the close-loop system. Our scheme is equivalent to use spectrally tunable environment to directly couple the controlled system. In this regard, our scheme is a direct coherent feedback scheme~\cite{LLoyd2000,XuePRA2012}. This similar loop topology has been employed to photonic crystal systems, by which the noise is driven out of resonance with the working frequency of the system so as to suppress non-Markovian decoherence. However, the circumstance for the quantum dots system is quite different due to the bias voltage applied on the source and drain baths. The resulting detuning between the central frequencies of the source and the drain leads to a memory kernel function that is nonlinearly dependent on the feedback coupling strength, which makes it difficult to design coherent feedback.

In this paper, we utilize a Green's function based root locus method~\cite{xueqip2015,xueian} to analyze the decoherence effect of the closed-loop system, by which we show that the coherent feedback can suppress the decoherence in quantum dots systems. The rest of this paper is organized as below. In section \ref{2nd}, the Hamiltonian of the coherent feedback loop is introduced. Starting from this Hamiltonian, we obtain an exact non-Markovian Langevin equation to describe the dynamics of the quantum dot in section \ref{3rd}. In section \ref{4th}, via Green's function based root locus approach, the analysis for the dynamics of the controlled system is done in the frequency domain. An example of the quantum dot system is given in section \ref{5th}. Finally, conclusions are drawn in section \ref{6th}.
\section{Coherent feedback loop Hamiltonian}\label{2nd}

\begin{figure}
  \includegraphics[width=8cm]{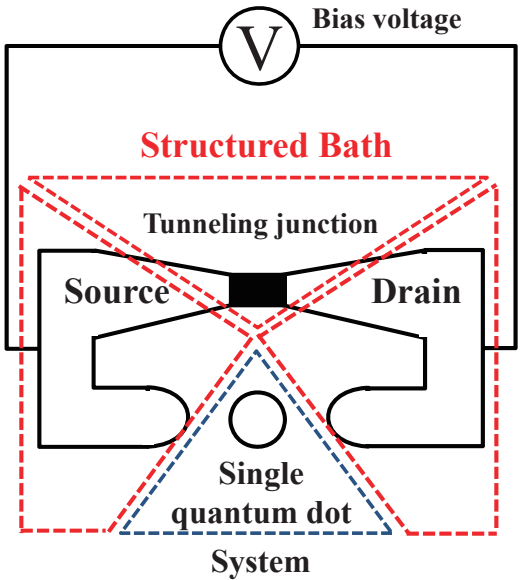}\\
  \caption{The schematic diagram of a direct coherent feedback loop for the quantum dot system.}\label{Fig1}
\end{figure}

Consider a single quantum dot~\cite{ReimRMP2002} located in the center of two leads named source (left) and drain (right), respectively, where a bias voltage is applied on the two leads. The coupling strengthes between the quantum dot and each mode of two electrodes are different resulting in the non-Markovian decoherence dynamics of the quantum dot~\cite{Tu2008}. To effectively reject non-Markovian noises, non-Markovian coherent feedback scheme is introduced. To build a feedback loop, the source and drain are joined together with a tunneling junction where the tunneling strength is tunable. This scheme is sketched in Fig.~\ref{Fig1}. This design leads to a closed interaction relationship, where the interconnection of each part induces bidirectional causal effect (i.e., two interconnected system always affect each other) . Thus, the information flow in this closed loop is in both clockwise and anticlockwise directions.

The Hamiltonian of the open-loop system (i.e., without tunneling junction) can be written as
\begin{equation}\label{1}
    H_O=H_{S}+H_E+H_{SE},
\end{equation}
where $H_{S}/\hbar=\omega_{S} \hat{d}^\dagger \hat{d}$ is the quantum dot Hamiltonian with a working frequency $\omega_S$ and a fermion annihilation operator $\hat{d}$. The environment Hamiltonian $H_E$ describes two clusters of the electron bath (source and drain), i.e., $$H_E/\hbar=\sum_k\omega_{Bk}\hat{b}^\dagger_{k} \hat{b}_{k}+\sum_k\omega_{Ck}\hat{c}^\dagger_{k} \hat{c}_{k}$$ with the frequencies for each mode $\omega_{Bk}$ ($\omega_{Ck}$), where the symbols $\hat{b}_{k}$ ($\hat{c}_{k}$) are the fermion annihilation operator of the source (drain). Their couplings to the system are determined by the interaction Hamiltonian $$H_{SE}/\hbar=\sum_k (V^*_{Bk}\hat{d}^\dagger \hat{b}_{k}+V_{Bk}\hat{b}^\dagger_{k}\hat{d})+\sum_k (V^*_{Ck}\hat{d}^\dagger \hat{c}_{k}+V_{Ck}\hat{c}^\dagger_{k}\hat{d}).$$ The coupling strengthes between the system and each mode of the source~(drain) are denoted as $V_{Bk}$($V_{Ck}$), which is different for each mode resulting in the non-Markovian decoherence dynamics. For simplicity, the nonlinear system-bath interactions are not considered here.

For the above system, the interaction between the system and the bath disturbs the system dynamics. The noise structure induced by the interaction determines how serious the non-Markovian decoherence is. In this paper, a tunnelling junction is introduced between the source and the drain to efficiently modify the noise structure, which induces a coupling Hamiltonian between the source and drain as
\begin{eqnarray}\label{2}
H_{F}&=&\sum_k\sum_{k'}(F_{kk'}\hat{c}^\dagger_{k'}\hat{b}_{k}+F_{kk'}^*\hat{b}_{k}^\dagger\hat{c}_{k'})
\end{eqnarray}
 where $F_{kk'}$ describes the tunneling strength between the $k$-th source mode and the $k'$-th drain mode~\cite{Mahan}. Here, the source together with the drain constitutes a structured bath for the system (as shown in Fig.\ref{Fig1}), whose internal properties are expected to be modified via the tunable coupling strength $F_{kk'}$.

The $F_{kk'}$ will depend on the physical properties of the junction. In what follow we describe how to calculate $F_{kk'}$, assuming the electrons act as if they are free and are in one dimension. Consider an electron starting at the source with wave vector $k_B$ and ending at the drain with wave vector $k_C$, an initial energy and a final energy can be expressed as $E_B = \hbar^2 k_B^2 / 2m +  eU_B$ and $E_C = \hbar^2 k_C/ 2m + eU_C$, respectively, where $U_B$ and $U_C$ are the voltage on each side of the junction, $e$ is the charge of the electron and $m$ is its mass.
By energy conservation we can relate the initial and final wave vector of the electron after crossing the junction as: $ \hbar^2(k_C^2 - k_B^2)/ 2m  = e(U_B - U_C)$.

On the other hand, we assume that the initial and final kinetic energy of the electron is much larger than the potential difference across the junction. In this case the probability is very low for the electron to be reflected.
Hence, the characteristic central frequency $k_0$ of both baths is much higher than the potential difference across the junction, i.e., $\hbar^2k_0^2/2m \gg e(U_C - U_B)$.
 Furthermore, we only consider perturbations about whose frequency components are near this central frequency, as the off-resonant components have minor effects on the coherence of the quantum dot. Thus, it is easy to see that $ k_C - k_B  \approx m e(U_B - U_C)/\hbar^2 k_0$.
Hence, the difference between the wave vectors is approximately a constant which is related to the potential difference across the junction. And thus the tunneling strength between the source and drain can be expressed as
\begin{equation}\label{2-0}
F_{kk'}=\left\{\begin{array}{cc}
          f_{k}, & k-k'=l\neq0 \\
          0, & {\rm otherwise}
        \end{array}\right.
\end{equation}
where $l = m e(U_B - U_C)/\hbar^2k_0$. The coupling strengths $f_{k}$ can be engineered by treating the junction as a waveguide and changing its geometry.

 Thus, the total Hamiltonian of our coherent feedback control system reads as
\begin{equation}\label{14}
H_{T}=H_O+H_{F}.
\end{equation}
The details of the rejection of the non-Markovian noises via coherent feedback will be shown in the next sections.

\section{Exact Non-Markovian Quantum Langevin Equation}\label{3rd}
\subsection{Exact Langevin equation}
The evolution of the fermion annihilation operator $\hat d(t)$ of the quantum dot is described by the following integral-differential exact non-Markovian quantum Langevin equation (for the details of the derivation, see the Appendix~\ref{ASA}),
\begin{equation}\label{6}
 \dot{\hat d}(t)=-i\omega_S\hat d(t)-\int_0^t  d\tau M(t-\tau)\hat d(\tau)-i\hat \e_n(t),
\end{equation}
where the memory kernel function $M(t)$ embedded with the noise spectrum determines the dissipation process; and the noise term $\hat \e_n(t)$ corresponds to the equivalent noise injected from the two leads.

Due to the linearity of the integral-differential Eq.~(\ref{6}), the solution of $\hat{d}(t)$ is expressed as
\begin{equation}\label{10}
\hat{d}(t)=g(t)\hat{d}(0)+\int_0^td\tau g(t-\tau)\hat \e_n(\tau),
\end{equation}
where the first term characterizes the dissipative evolution from its initial state $\hat{d}(0)$ and the second term describes the dynamics excited by the noise $\hat\e_n(t)$. The complex coefficient $g(t)$ satisfies the following integral-differential equation:
\begin{equation}\label{11}
\dot{g}(t) = -i\omega_S g(t)- \int_0^t  d\tau M(t-\tau)g(\tau),\quad g(0)=1,
\end{equation}
where the absolute value of the Green's function $g(t)$ is the scaled amplitude of the system~\cite{Tan2011}. It can be used to evaluate the dissipation process of the system due to affected by the same memory kernel function $M(t)$ as that for the system operator $\hat{d}(t)$.
\subsection{The coherent feedback case}
When the feedback couplings (\ref{2}) are introduced, i.e., the total system is described by (\ref{14}), both $M(t)$ and $\hat \e_{\rm n}(t)$ are affected. Assume that the source and drain can be effectively coupled as expressed in Eq.~(\ref{A1}) and denote $F_{kk'}$ in the continuous limit and the polar coordinate as $f(\omega)=r(\omega)e^{ i\theta(\omega)}$. The memory kernel function is split by the feedback as $M(t)\equiv M_f(t)=M^+(t)+M^-(t)$ with
\begin{equation}\label{12}
M^\pm(t) =\int_{-\infty}^{+\infty}d\omega \frac{J^\pm(\omega)}{2\pi} e^{-i(\omega-\delta\pm\sqrt{\delta^2+r(\omega)^2})t},
\end{equation}
where the noise spectral functions
\begin{eqnarray}
\frac{J^+(\omega)}{2\pi} &=&\varrho(\omega) |V_{B}(\omega)e^{- i\theta(\omega)}\cos\frac{\alpha(\omega)}{2}+V_{C}(\omega)\sin\frac{\alpha(\omega)}{2}|^2,\nonumber \\
\frac{J^-(\omega)}{2\pi} &=&\varrho(\omega)|V_{B}(\omega)e^{- i\theta(\omega)}\sin\frac{\alpha(\omega)}{2}-V_{C}(\omega)\cos\frac{\alpha(\omega)}{2}|^2\nonumber
\end{eqnarray}
are modulated by feedback parameters $r(\omega)$ and $\theta(\omega)$ and $\alpha(\omega)=\arctan\frac{r(\omega)}{\delta}$. The split in the memory kernel function shows that the noises can be modified by the tunneling strength. The equivalent noise $\hat \e_n(t)\equiv\hat \e_{nf}(t)$ in Eq.~(\ref{6}) is
\begin{equation}
\hat \e_ {nf}(t)=\int_{-\infty}^{+\infty} d\omega \varrho(\omega){v}^\dagger(\omega)\Phi(\omega,t){\hat \e}(\omega,0),
\end{equation}
where $\varrho(\omega)$ is the density state and the definitions of the coupling strength vector $v(\omega)$ and the feedback-induced modulation matrix $\Phi(\omega,t)$ are given in Eqs.~(\ref{13-55}) and (\ref{51}).
Note that we have assumed the source and drain share the same density state $\varrho(\omega)$ and the effect of the feedback Hamiltonian $H_F$ has been embedded in both Green's function $g(t)$ and the equivalent input $\hat \e_n(t)$.
\subsection{The open-loop case}\label{opl}
For comparing with an open-loop method to suppress the non-Markovian decoherence~\cite{Lei2011}, the memory kernel function and the noise term in the open-loop case are also considered.
In absence of feedback couplings (\ref{2}), i.e., the system is totally described by the open-loop Hamiltonian $H_O$ (\ref{1}), the memory kernel function $M(t)\equiv M_0(t)=M_B(t)+M_C(t)$ with
\begin{eqnarray}\label{27}
M_B(t)&=& \int_{-\infty}^{+\infty} d\omega_B \varrho(\omega_B)|V_B(\omega_B)|^2 e^{- i\omega_{B}t},\\
M_C(t)&=&\int_{-\infty}^{+\infty} d\omega_C \varrho(\omega_C)|V_C(\omega_C)|^2 e^{- i\omega_{C}t},
\end{eqnarray}
is only dependent on the system coupling strengthes with the source and drain, where $V_{B}(\omega_{B})$ and $V_{C}(\omega_{C})$ are the coupling strength of the system with the source and drain in a continuous frequency form, respectively. And $\varrho (\omega_{B})$ and $\varrho (\omega_{C})$ are the state density function of the source and drain, respectively;
and the noise $\hat \e_{n}(t)\equiv\hat \e_{n0}(t)=\hat \e_{\rm B0}(t)+\hat \e_{\rm C0}(t)$
is a summation of the noise arising from the source and drain
\begin{eqnarray}
\hat \e_{\rm B0}(t)&=&\int_{-\infty}^{+\infty} d\omega_{B}\varrho (\omega_{B}) V^*_{B}(\omega_{B})e^{-i\omega_{B}t}\hat{b}(\omega_{B},0)\\
\hat \e_{\rm C0}(t)&=&\int_{-\infty}^{+\infty}d\omega_{C}\varrho (\omega_{C}) V^*_{C}(\omega_{C}) e^{- i\omega_{C}t}\hat{c}(\omega_{C},0).
\end{eqnarray} In the above expression,
$\hat{b}(\omega_{B},0)$ and $\hat{c}(\omega_{C},0)$ are the value of $\hat b(\omega_{B},t)$, $\hat c(\omega_{C},t)$ at $t=0$, respectively.



The exact non-Markovian Langevin equation above affords the basis for analyzing the system dynamics with or without coherent feedback control. The feedback control parameters $r(\omega)$ and $\theta(\omega)$ are embodied in the memory kernel function $M(t)$ in Eq.~(\ref{6}). How to effectively manipulate the memory kernel function $M(t)$ is considered in the next section.

\section{Green's function based Root locus Analysis for decoherence suppression}\label{4th}

In our previous work\cite{XuePRA2012}, it is shown that spectral modulation induced by coherent feedback can be used to suppress decoherence. However, this method is not directly extendable to the system discussed here due to the nonlinearity of the control amplitude $r(\omega)$ as shown in Eq.~(\ref{12}) resulting from the bias-voltage-induced central frequency difference between the source and drain. Hence, whether or not decoherence can be suppressed is not as obvious as in Ref.~\onlinecite{XuePRA2012}. In this section, we will analyze it through Green's function based root locus method.

\subsection{Green's function based root locus}
Root locus is a graphical method for describing the dependence of the modes on a changeable parameter of the controlled system (e.g., the gain) and thus determining the regime of the parameter that ensures the system stability~\cite{Ogata}. Here, we analyze the root locus for the Green's function to understand the mechanism of decoherence suppression induced by coherent feedback.

Transforming the dynamical equation of the Green's function for the non-Markovian quantum system~(\ref{11}) into the complex frequency domain, the Laplace transform $G(s)$ of the Green's function $g(t)$ is
\begin{equation}\label{15}
G(s)=\frac{1}{s+ i\omega_S+M(s)}~.
\end{equation}
where $M(s)$ is the Laplace transform of the memory kernel function $M(t)$. The poles of the Green's function $G(s)$ are defined as points of $s$ at which $G(s)$ is singular. The trajectories of the poles versus a varying parameter as the root locus of the Green's function $G(s)$~\cite{xueqip2015}.

As shown in Eq.~(\ref{15}), the poles of the Green's function are dependent with the memory kernel function $M(s)$. For the simplest case $M(s)=0$, i.e., the system is closed, the pole lies in the imaginary axis of the complex plane, which implies the coherence of the system are not destroyed. When the system is a Markovian quantum system, i.e., $M(s)=\frac{\gamma}{2}$ with a constant damping rate $\gamma$, the pole is shifted to the left half of the complex plane with a negative real part corresponding to the damping. For a non-Markovian quantum system involving complicated noise spectrum, the distribution of the poles of the Green's function becomes complicated, where we assume that $M(s)$ can be expressed in a rational form. In the following, we will investigate its influence of the memory kernel function on the Green's function in the case with or without our coherent feedback scheme, respectively, so as to observe the coherence of the system.

\subsection{The coherent feedback case}
To explore the root locus of the Green's function induced by the coherent feedback, we assume that the single quantum dot is equally strongly coupled to the source and drain, i.e., $V_B(\omega)=V_C(\omega+2\delta)=V(\omega)$ where $2\delta$ is the frequency difference between the two baths, and the Lorentzian spectral density~\cite{Tu2008,WeiMinPRL2012} is adopted here for fermion systems as
\begin{equation}\label{18}
J(\omega)=2\pi\varrho(\omega)|V(\omega)|^2=\frac{\eta h^2}{(\omega-\omega_S)^2+h^2}~,
\end{equation}
where the parameters $\eta$ and $h$ are the strength and width of the noise spectrum, respectively.

Assume that the feedback coupling strength is independent on the frequency and express it in the polar coordinate as $f=re^{ i\theta}$, and thus the corresponding parameter $\alpha$ is also independent on frequency. When the feedback coupling is applied, the memory kernel function $M(t)$ is split into two branches in Eq.~(\ref{12}) which can be expressed as $M(s)=M^+(s)+M^-(s)$ in the frequency domain (see Appendix~\ref{ASC}) with
\begin{equation}\label{19}
M^\pm(s)=\frac{\frac{1}{2}\eta h(1\pm\cos\theta\sin\alpha)}{s+z_0\pm i\gamma},
\end{equation}
where $z_0=h+i(\omega_S-\delta)$ and $\gamma=\sqrt{\delta^2+r^2}$. Physically, this means our coherent feedback can modify the noise spectrum, i.e., the structure of the environment can be engineered by the coherent feedback.

To see how the memory kernel $M(s)$ affects the Green's function, we can substitute Eq.~(\ref{19}) into Eq.~(\ref{15}) and then obtain
\begin{equation}\label{21}
G(s)=\frac{s^2+\alpha_1s+\alpha_2}{s^3+\beta_1s^2+\beta_2s+\beta_3},
\end{equation}
where $\alpha_1=2z_0$,
$\alpha_2=z_0^2+\gamma^2$,
$\beta_1=2z_0+i\omega_S$,
$\beta_2=z_0^2+\gamma^2+\eta h+i2\omega_S z_0$, and
$\beta_3=\eta hz_0+i\omega_S(z_0^2+\gamma^2)-i\eta h\gamma\cos\theta\sin\alpha$.

For utilizing inverse Laplace transform to obtain an explicit solution of $g(t)$ in the time domain, we express Eq.~(\ref{21}) in the form of a partial fraction decomposition as
\begin{equation}\label{22}
G(s)=\frac{q_1}{s-p_1}+\frac{q_2}{s-p_2}+\frac{q_3}{s-p_3}~,
\end{equation}
where three poles
\begin{eqnarray}
p_1&=&-\frac{\beta_1}{3}+\frac{l}{3\sqrt[3]{2}}e^{i\phi}-\frac{\sqrt[3]{2}A}{3l}e^{-i\phi},\\
p_2&=&-\frac{\beta_1}{3}-\frac{l}{3\sqrt[3]{2}}e^{i(\phi-\frac{\pi}{3})}+\frac{\sqrt[3]{2}A}{3l}e^{-i(\phi-\frac{\pi}{3})},\\
p_3&=&-\frac{\beta_1}{3}-\frac{l}{3\sqrt[3]{2}}e^{i(\phi+\frac{\pi}{3})}+\frac{\sqrt[3]{2}A}{3l} e^{-i(\phi+\frac{\pi}{3})},
\end{eqnarray}
 with $
A=3\beta_2-\beta_1^2,
B=9(\beta_2\beta_1-3\beta_3)-2\beta_1^3
$,
and
$le^{i\phi}\equiv\sqrt[3]{B+\sqrt{4A^3+B^2}}$ are what we concern about. Their distribution determines the root locus of the Green's function and thus the non-Markovian dynamics of the system.
The complex coefficients $q_1,q_2,q_3$ can be calculated as
\begin{eqnarray}\label{23}
q_1&=& \frac{\alpha_2 + \alpha_1 p_1 + p_1^2}{(p_1 - p_2) (p_1 - p_3)},\nonumber\\
q_2&=& \frac{-\alpha_2 - \alpha_1 p_2 -  p_2^2}{(p_1 - p_2) (p_2 - p_3)},\nonumber\\
q_3&=&\frac{\alpha_2 + \alpha_1 p_3 + p_3^2}{(p_1 - p_3) (p_2 - p_3)}.\nonumber
   \end{eqnarray}
With the help of Eq.~(\ref{22}), the solution of Eq.~(\ref{11}) can be obtained as
\begin{eqnarray}\label{20}
g(t)=q_1 e^{p_1 t}+q_2 e^{p_2 t}+q_3 e^{p_3 t},
\end{eqnarray}
which will be used to observe the dynamics of $g(t)$ under coherent feedback in the example of next section.

The number of poles of $G(s)$ is increased to be 3. Their distribution will directly affect the dynamics of $g(t)$. To qualitatively observe the effect of our coherent feedback on the distribution of the poles of the Green's function $G(s)$, we consider a limit case that the feedback coupling strength $r$ approaches to infinity. Since the three poles $p_{1,2,3}$ are functions of the feedback coupling strength $r$, the limit value of $p_{1,2,3}$ as $r$ going to the infinity are calculated as
\begin{eqnarray}
\lim_{r\rightarrow+\infty}p_1&=&0+ i(-\omega_S),\\
\lim_{r\rightarrow+\infty}p_2&=&-h+ i(-\infty),\\
\lim_{r\rightarrow+\infty}p_3&=&-h+ i(+\infty).
\end{eqnarray}
It is shown that one of the poles $p_1$ is pushed to be close to $-i\omega_S$ by choosing a sufficiently large $r$ and the real value of the other two poles are driven to be $-h$ and the imaginary parts of them go to $-\infty$ and $+\infty$, respectively. Compared with $p_2$ and $p_3$ whose real parts are negative leading to quick damping, the pole $p_1$ for sufficiently large $r$ is very close to the imaginary axis, which will keep its mode oscillating for a long time. It means $|g(t)|$ can be kept on a high value close to $1$. This indicates our coherent feedback scheme can suppress the decoherence. In practice, the feedback coupling strength can not be arbitrarily strong, and hence the dissipation process can only be slowed down only by the feedback.
\subsection{The open-loop case}

If the Hamiltonian $H_{F}$ is ignored, our system is reduced to a common single quantum dot setting governed by the open loop Hamiltonian $H_O$. The system dynamics obeys the same form Langevin equation~(\ref{6}) with a different memory kernel $M_0(t)$ and a noise term $\hat \e_{n0}(t)$ as given in section~\ref{opl}.

Transformed to the frequency domain, (see the Appendix~\ref{ASC}), $M_0(s)$ in Eq. (\ref{15}) is expressed as
\begin{eqnarray}\label{31}
M_0(s)=\frac{\frac{1}{2}\eta h}{s+h+ i\omega_S}+\frac{\frac{1}{2}\eta h}{s+h+i(\omega_S-2\delta)}.
\end{eqnarray}
Substituting Eq.~(\ref{31}) into Eq.~(\ref{15}), a partial fraction decomposition form of $G_0(s)$ can be obtained as
\begin{equation}\label{25}
G_0(s)=\frac{q_{01}}{s-p_{01}}+\frac{q_{02}}{s-p_{02}}+\frac{q_{03}}{s-p_{03}},
\end{equation}
where three poles $p_{01}$, $p_{02}$, $p_{03}$ of $G_0(s)$ are equal to the values of $p_{1}$, $p_{2}$, $p_{3}$ as $r$ and $\theta$ being zero; and $q_{01},q_{02},q_{03}$ can also obtained in the same way.
Hence, the behavior of the Green's function $g_0(t)$ can be evaluated by
\begin{eqnarray}\label{24}
g_0(t)=q_{01} e^{p_{01} t}+q_{02} e^{p_{02} t}+q_{03} e^{p_{03} t},
\end{eqnarray}
which can be obtained from Eq.~(\ref{25}) via inverse Laplace transform.

Ref.~\onlinecite{Lei2011} proposed a scheme of realizing strong couplings between the system and its environment to suppress non-Markovian decoherence for bosonic systems. With respect to our system, it is equivalent to increase the noise strength $\eta$ in Eq.~(\ref{18}) to suppress the decoherence. In the next section, we will numerically compare the method in Ref.~\onlinecite{Lei2011} with our coherent feedback scheme.

\section{Example of single quantum dot}\label{5th}
In numerical simulations, we choose parameters that can be engineered as follows: the system working frequency $\hbar\omega_S=10\mu eV$, the frequency difference between the source and drain $\hbar\delta=0.05\mu eV$and the noise width $\hbar h=0.3\mu eV$. Other varying parameters will be given below. The coherence of the system is measured by absolute value of the Green's function $|g(t)|$ which can be analytically calculated as Eq.~(\ref{20}) or Eq.~(\ref{24}).
\begin{figure}
  \includegraphics[width=8.5cm]{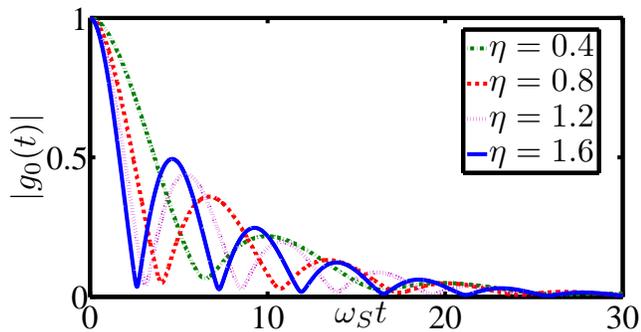}\\
  \caption{(Color online) The dynamics of the absolute value of the Green's function $|g_0(t)|$ with increasing the noise strength $\eta$. The decay of $|g_0(t)|$ is not apparently improved.}\label{Fig5}
\end{figure}

Figure~\ref{Fig5} shows the variations of the absolute value of open-loop Green's function $g_0(t)$ with increasing the noise strength to realize strong couplings between the system and the bath. When the noise strength $\eta$ is set to be $0.4$, $|g_0(t)|$ is oscillatingly damping as plotted in green dot-dashed line, which indicates that the dynamics of the system is in the non-Markovian regime. When $\eta$ is further increased, e.g., $\eta=0.8$ or $\eta=1.2$, the oscillation of $|g_0(t)|$ is enhanced. However, the damping of $|g_0(t)|$ can not be stopped. Even when $\eta$ reaches $1.6$, the damping of $|g_0(t)|$ is still not changed. As pointed out by Ref.~\onlinecite{Lei2011}, the damping process can be slowed down via increasing the coupling strength between the system and the baths (equivalent to increase the noise strength $\eta$) for boson systems. Here, we observe that their strategy does not work for fermion systems.

\begin{figure}
  \includegraphics[width=8.5cm]{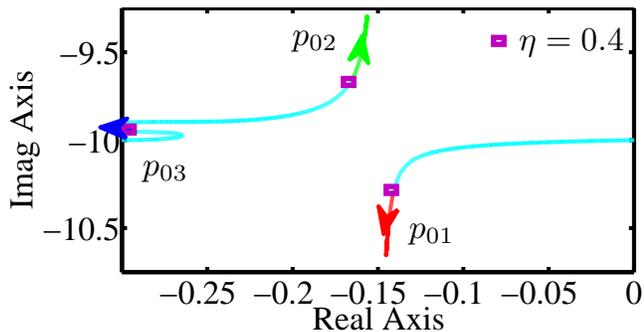}\\
  \caption{(Color online) The poles variation of the open-loop Green's function $G_0(s)$ versus increasing noise strength $\eta$ from $0$ to $1.6$. The variation of the poles $p_{01}$, $p_{02}$, $p_{03}$ in the non-Markovian regime corresponding to the Figure~\ref{Fig5} are plotted in red, green, and blue lines, respectively. The starting points $\eta=0.4$ are labeled by the rectangle mark. This figure shows the scheme increasing noise strength as done in Ref.~\onlinecite{Lei2011} can not effectively drive the poles to be close to the imaginary axis so as to slow down the damping.}\label{Fig2}
\end{figure}

The reason can be understood in the root locus plot Figure~\ref{Fig2} which shows the dependence of three poles of the open-loop Green's function $g_0(t)$ with the noise coupling strength $\eta$ from $0.4$ to $1.6$. It is clearly shown $p_{01}$ and $p_{02}$ are towards the axis ${\rm Re}s=0.15$ with opposite directions and $p_{03}$ is away from the imaginary axis. The three modes of $g_0(t)$ have negative real parts no matter how large $\eta$ is. The strategy in Ref.~\onlinecite{Lei2011} can not decrease the real part value of poles approaching to the imaginary axis, which indicates the damping process can not be suppressed via increasing the couplings between the system and the bath.

\begin{figure}
  \includegraphics[width=8.5cm]{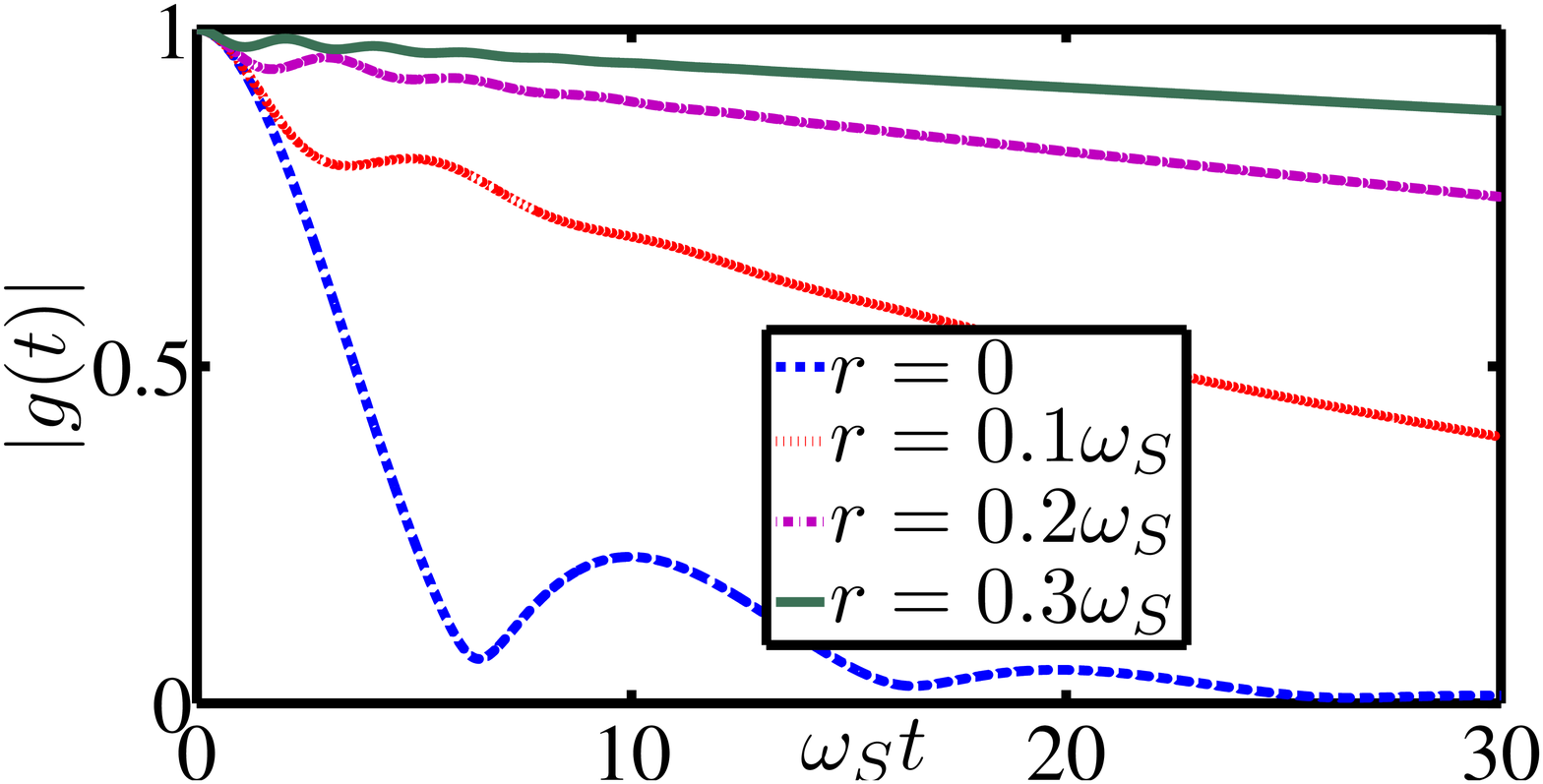}\\
  \caption{(Color online)The dynamics of the absolute value of the Green's function $|g(t)|$ versus the increasing feedback coupling strength $r(\omega)=r$. With the increasing $r$, the value of $|g(t)|$ is kept on a high value for a long time.}\label{Fig4}
\end{figure}

Compared with the open-loop case, the decoherence can be significantly suppressed via coherent feedback as shown in Figure~\ref{Fig4}, where the noise strength is set sufficiently large, e.g., $\eta=0.4$ causing the non-Markovian dynamics of the system (see the blue dashed line under $r=0$ in Figure~\ref{Fig4}). When the feedback loop is closed, e.g., $r=0.1\omega_S$, the damping of the absolute value of $g(t)$ is slowed down. The value of $|g(t)|$ can be kept on a high value when the feedback strength is further enhanced, for example, $r=0.2\omega_S$ or $r=0.3\omega_S$.

\begin{figure}
  \includegraphics[width=8.5cm]{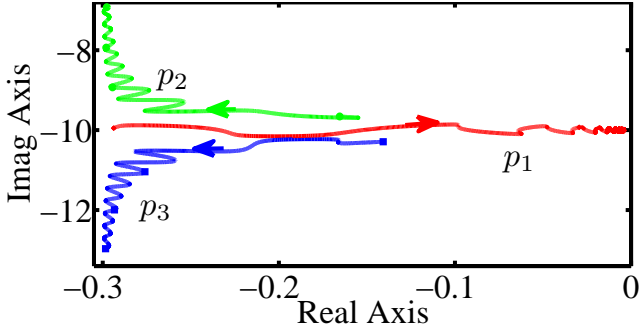}\\
  \caption{(Color online) The root locus  of the Green's function $G(s)$ with respect to the feedback coupling strength $r(\omega)=r$ from $0$ to $0.3\omega_S$ with the noise strength $\eta=0.4$. The pole $p_1$ is oscillatingly pushed to be close to the imaginary axis so as to afford a very slow damping mode of $g(t)$, which indicates the decoherence is effectively suppressed by our coherent feedback scheme.}\label{Fig3}
\end{figure}

The above phenomena can be well analyzed from the variation of poles of $g(t)$ with continuously increasing the feedback coupling strength $r$ from $0$ to $0.3\omega_S$ (see root locus plot Figure~\ref{Fig3}). The three poles initially lie in the left-part of complex plane with negative real parts (as shown the starting point of three lines) causing the damping of $|g(t)|$. When the feedback coupling strength is enhanced, the pole $p_1$ is oscillatingly driven to be close to the imaginary axis and the other poles $p_2,p_3$ are pushed to approach to ${\rm Re}s=-0.3$. We can see that the real part of the pole $p_1$ is nearly decreased to be zero when the feedback coupling strength is sufficiently strong, which indicates that such weak damping mode can help $g(t)$ to resist the decoherence. Compared with the open loop strategy above, our coherent feedback scheme can effectively suppress the decoherence in quantum dot systems.

\section{Conclusion}\label{6th}
This paper presents a non-Markovian coherent feedback scheme to stabilize a single quantum dot whose natural noise baths (source and drain) are connected to form the tunable quantum tunneling process. The mechanism of the decoherence suppression is analyzed in the frequency domain via the root locus of the Green's function which is extended from classical control theory. Compared with the open loop strong coupling strategy, our coherent feedback scheme can suppress the damping of the system dynamics more efficiently.

For future works, it is worthwhile to explore how to apply our direct coherent feedback scheme to complicated quantum dots systems, e.g., two quantum dots system where the Coulomb interaction between two dots exists. In addition, when a quantum dot is weakly coupled with a resonator, its information can be indirectly extracted through the output of a probing field of the resonator~\cite{PhysRevLett.108.046807}. Hence, this makes it possible to design a field-mediated coherent feedback controller for the quantum dot.
\appendix
\section{Derivation of non-Markovian Langevin Equation under feedback}\label{ASA}
To facilitate the following derivation, we assume
\begin{equation}\label{A1}
F_{kk'}=\left\{\begin{array}{cc}
          f_{k} & k-k'=l\neq0 \\
          0 & {\rm otherwise}
        \end{array}\right.
\end{equation}
which implies only two modes with mode difference $l$ in each bath can be effectively coupled.

According to the Heisenberg equation in quantum mechanics
\begin{equation}\label{3}
    \dot{\hat{o}}(t)=-\frac{i}{\hbar}[\hat{o}(t),H(t)]
\end{equation}
for arbitrary operator $\hat{o}(t)$. The motion equations for the system and bath modes can be obtained as
\begin{eqnarray}\label{4}
\dot{\hat{d}}(t)&=&- i\omega_S \hat{d}(t)- i\sum_kV^*_{Bk}\hat{b}_{k}(t)- i\sum_{k-l}V^*_{Ck-l}\hat{c}_{k-l}(t),\nonumber\\
&&\label{4-1}\\
\dot{\hat{b}}_{k}(t)&=&-i\omega_{Bk}\hat{b}_{k}(t)- if_k^*\hat{c}_{k-l}(t) - i V_{Bk}\hat{d}(t),\label{4-2}\\
\dot{\hat{c}}_{k-l}(t)&=&-i\omega_{Ck-l}\hat{c}_{k-l}(t)- if_k \hat{b}_{k}(t)- i V_{Ck-l}\hat{d}(t).\label{4-3}
\end{eqnarray}
Firstly, the motion coupling equations between two bath (\ref{4-2}) and (\ref{4-3}) can be jointly solved as
\begin{equation}\label{5}
   \hat{\e}_{k}(t)=\Phi_{k}(t)\hat{\e}_{k}(0)- i\int_0^t \Phi_{k}(t-\tau)v_{k}\hat{d}(\tau) d\tau,
\end{equation}
where
$$\hat{\e}_{k}(t)=\left[
                      \begin{array}{c}
                        \hat{b}_{k}(t) \\
                        \hat{c}_{k-l}(t) \\
                      \end{array}
                    \right]
, v_{k}=\left[
            \begin{array}{c}
              V_{Bk} \\
              V_{Ck-l} \\
            \end{array}
          \right].
$$
Expressing the feedback coupling strength in the polar coordinate as $f_{k}=r_{k}e^{i\theta_{k}}$, the transition matrix is calculated as
\begin{eqnarray}\label{51}
& &\Phi_{k}(t)= \exp\left[- it\left(
    \begin{array}{cc}
                                        \omega_{Bk} & f_{k}^* \\
                                      f_{k}  &  \omega_{Ck-l} \\
                                         \end{array}
   \right)\right]\nonumber\\
   &&=\left[
     \begin{array}{cc}
       \chi_+e^{-i\lambda_+t}-\chi_- e^{- i\lambda_-t} &\kappa^*(e^{- i\lambda_+t}-e^{-i\lambda_-t})  \\
       \kappa(e^{-i\lambda_+t}-e^{- i\lambda_-t})  & - \chi_- e^{-i\lambda_+t}+\chi_+ e^{- i\lambda_-t} \\
     \end{array}
   \right],\nonumber\\
 \end{eqnarray}
where $\chi_\pm=\frac{1}{2}(\cos\alpha_k\pm 1)$ and $\kappa=\frac{1}{2}\sin\alpha_k e^{i\theta_k}$ with $\alpha_k=\arctan\frac{r_k}{\delta}$ and frequency difference $2\delta=\omega_{Bk}-\omega_{Ck-l}$. Eigenvalues $\lambda_\pm$ of the matrix $\left[
    \begin{array}{cc}
                                        \omega_{Bk} & f_{k}^* \\
                                      f_{k}  &  \omega_{Ck-l} \\
                                         \end{array}
   \right]$
are expressed as
\begin{equation}
\lambda_\pm=\frac{\omega_{Bk}+\omega_{Ck-l}\pm\sqrt{(\omega_{Bk}-\omega_{Ck-l})^2+4r_{k}^2}}{2}.
 \end{equation}

Then, substituting (\ref{5}) into (\ref{4-1}), we can get the system Langevin equation as
\begin{equation}\label{7}
\dot{\hat d}(t)=-i\omega_{S} \hat{d}(t)-\int_0^t d\tau M(t-\tau) \hat d(\tau)-i\hat \e_{\rm n}(t),
\end{equation}
where the memory kernel function and the equivalent noise are defined as
\begin{equation}\label{8}
 M(t) = \sum_k v_{k}^\dagger \Phi_{k}( t) v_{k},\quad \hat \e_{\rm n}(t)=\sum_k v_{k}^\dagger\Phi_{k}(t)\hat{\e}_{k}(0),
\end{equation}
respectively.

The memory kernel function $M(t)$ can be further expressed as
\begin{eqnarray}\label{9}
M(t)& = &\sum_k |V_{Bk}e^{-i\theta_k}\cos\frac{\alpha_k}{2}+V_{Ck-l}\sin\frac{\alpha_k}{2}|^2e^{-i\lambda_+t}\nonumber\\
&&+\sum_k |V_{Bk}e^{-i\theta_k}\sin\frac{\alpha_k}{2}-V_{Ck-l}\cos\frac{\alpha_k}{2}|^2e^{-i\lambda_-t}\nonumber\\
\end{eqnarray}
which are modulated by $r_k$ and $\theta_k$.

Under the continuous limit that the modes of the baths are too dense, the memory kernel function $M(t)$ is expressed in a frequency continuous form and can be further decomposed as $M(t) =M^+(t)+M^-(t)$ with
\begin{equation}\label{A12}
M^\pm(t) =\frac{1}{2\pi}\int_{-\infty}^{+\infty} d\omega J^\pm(\omega)e^{-i\lambda_\pm(\omega)t}
\end{equation}
where the noise spectral functions are
\begin{equation}
\frac{J^+(\omega)}{2\pi}=\varrho(\omega) |V_{B}(\omega)e^{-i\theta(\omega)}\cos\frac{\alpha(\omega)}{2}+V_{C}(\omega)\sin\frac{\alpha(\omega)}{2}|^2\\
\end{equation}
\begin{equation}
\frac{J^-(\omega)}{2\pi}=\varrho(\omega)|V_{B}(\omega)e^{-i\theta(\omega)}\sin\frac{\alpha(\omega)}{2}-V_{C}(\omega)\cos\frac{\alpha(\omega)}{2}|^2,
\end{equation}
and
\begin{equation}\label{13}
    \lambda_\pm(\omega)=\omega-\delta\pm\sqrt{\delta^2+r(\omega)^2}.
\end{equation}
with the state density $\varrho(\omega)$. The noise term $ \hat{\e}_{\rm n}(t)$ in Eq.~(\ref{6}) is
\begin{equation}
    \hat{\e}_{\rm n}(t)=\int_{-\infty}^{+\infty}d\omega\varrho(\omega)v^\dagger(\omega)\Phi(\omega,t)\hat{\e}(\omega,0),
\end{equation}
where
\begin{equation}\label{13-55}
\hat{\e}(\omega,t)=\left[
                      \begin{array}{c}
                        \hat{b}(\omega,t) \\
                        \hat{c}(\omega-2\delta,t) \\
                      \end{array}
                    \right]
, v(\omega)=\left[
            \begin{array}{c}
              V_{B}(\omega,t) \\
              V_{C} (\omega-2\delta,t)\\
            \end{array}
          \right].
\end{equation}
and
\begin{eqnarray}\label{51}
&&\Phi(\omega,t)= \\
&&{\tiny \left[
     \begin{array}{cc}
       \chi_+e^{-i\lambda_+(\omega)t}-\chi_- e^{- i\lambda_-(\omega)t} &\kappa^*(e^{- i\lambda_+(\omega)t}-e^{-i\lambda_-(\omega)t})  \\
       \kappa(e^{-i\lambda_+(\omega)t}-e^{- i\lambda_-(\omega)t})  & - \chi_- e^{-i\lambda_+(\omega)t}+\chi_+ e^{- i\lambda_-(\omega)t} \\
     \end{array}
   \right]},\nonumber
 \end{eqnarray}
where $\chi_\pm=\frac{1}{2}(\cos\alpha(\omega)\pm 1)$ and $\kappa=\frac{1}{2}e^{i\theta(\omega)}\sin\alpha(\omega)$ with $\alpha(\omega)=\arctan\frac{r(\omega)}{\delta}$ and frequency difference $2\delta=\omega_{B}(\omega)-\omega_{C}(\omega-2\delta)$.

%

\section{Expression of Memory Kernel Function $M(t)$ in the Frequency Domain}\label{ASC}
Inserting the Lorentzian spectral density (\ref{18}) into Eq.~(\ref{27}), we obtain
\begin{eqnarray}\label{52}
M_0(t)&=&\frac{1}{2\pi}\int_{-\infty}^{+\infty} \frac{\eta h^2}{h^2+(\omega-\omega_S)^2} e^{-i\omega t}   d\omega+\nonumber\\
&&\frac{1}{2\pi}\int_{-\infty}^{+\infty} \frac{\eta h^2}{h^2+(\omega-\omega_S+2\delta)^2} e^{- i\omega t}   d\omega
\end{eqnarray}
The integration of Eq.~(\ref{52}) can be solved as
\begin{equation}
M_0(t)=\frac{1}{2}\eta h e^{-h|t|-i\omega_S t}+\frac{1}{2}\eta h e^{-h|t|- i(\omega_S-2\delta) t}.
\end{equation}
Further, via Laplace transform, we can obtain $M_0(s)$ as
\begin{equation}
M_0(s)=\frac{\frac{1}{2}\eta h}{s+h+ i\omega_S}+\frac{\frac{1}{2}\eta h}{s+h+ i(\omega_S-2\delta)}
\end{equation}
Following the same idea, we also assume $\omega_S\pm\sqrt{\delta^2+r^2}$ are much larger than the noise width $h$. Therefore, we directly write transformed $M^\pm(t)$ as
\begin{equation}
M^\pm(s)=\frac{\frac{1}{2}\eta h(1\pm\cos\theta\sin\alpha)}{s+h+ i(\omega_S-\delta\pm\sqrt{\delta^2+r^2})}
\end{equation}
\section{Definition of Laplace Transform}\label{LP}
Laplace and its inverse transform for arbitrary operator $\hat{o}(t)$ are defined as
\begin{eqnarray}
  \hat{O}(s) &=& \int_0^\infty \hat{o}(t) e^{-st} dt, \\
  \hat{o}(t) &=& \frac{1}{2\pi i}\int_{\sigma-i\infty}^{\sigma+i\infty}\hat{O}(s)e^{st}ds ,
\end{eqnarray}
respectively, where $s=\sigma+i\omega$.

\end{document}